\documentclass[namedreferences]{solarphysics}
\usepackage[optionalrh,solaenum]{spr-sola-addons} 
\usepackage{graphicx}                    
\usepackage{color}                       
\usepackage{url}                         

\newcommand{\aap}{{\it Astron. Astrophys. }}

\newcommand{\solphys}{{\it Solar Phys. }}
\newcommand{\apj}{{\it Astophys. J. }}

\begin{document}

\begin{article}

\begin{opening}

\title{Estimation of solar prominence magnetic fields based on the reconstructed 3D trajectories of prominence knots }

\author{Maciej~\surname{Zapi{\'o}r$^{1}$\sep
        Pawe{\l}~\surname{Rudawy}$^{1,2}$}
       }

\runningauthor{M.~Zapi{\'o}r,  P.~Rudawy}
\runningtitle{Estimation of solar prominence magnetic fields}

   \institute{$^{1}$ Astronomical Institute of the University of Wroc{\l}aw \\ 51-622 Wroc{\l}aw, ul. Kopernika 11, Poland\\
                     email: \url{zapior@astro.uni.wroc.pl}, \url{rudawy@astro.uni.wroc.pl}\\
   $^{2}$ Space Research Centre, Polish Academy of Sciences \\51-622 Wroc{\l}aw, ul. Kopernika 11, Poland\\
             }

\begin{abstract}
We present an estimation of the lower limits of local magnetic fields in quiescent, activated, and active (surges) promineces, based on reconstructed 3-dimensional (3D) trajectories of individual prominence knots. The 3D trajectories, velocities, tangential and centripetal accelerations of the knots were reconstructed using observational data collected with a single ground-based telescope equipped with a \textit{Multi-channel Subtractive Double Pass} imaging spectrograph. Lower limits of magnetic fields channeling observed plasma flows were estimated under assumption of the equipartition principle. Assuming approximate electron densities of the plasma $n_e = 5\times 10^{11}$ cm$^{-3}$ in surges and $n_e = 5\times 10^{10}$ cm$^{-3}$ in quiescent/activated prominences, we found that the magnetic fields channeling two observed surges range from 16 to 40 Gauss, while in quiescent and activated prominences they were less than 10 Gauss. Our results are consistent with previous detections of weak local magnetic fields in the solar prominences.
\end{abstract}

\end{opening}

 \section{Introduction}

Solar quiescent prominences observed in the hydrogen H$\alpha$ line ($\lambda$=6562.8~\AA{}) are long living and globally stable structures, having a typical density of the order of $10^{-13}$ g cm$^{-3}$ ($n_{e}$=10$^{10}$-10$^{11}$ cm$^{-3}$) and a typical temperature of the order of 10$^{4}$ K \cite{W1993,TH1995,H2010}. Prominence plasma is coupled to magnetic fields, which channel plasma flows, carry plasma away and/or support it against solar gravity \cite{AD2003,BC2010}. Magnetic fields insulate also relatively cold and dense prominence plasma from the surrounding hot and much more dilute coronal plasma. Although the overall structure of the quiescent prominence is relatively stable, numerous small-scale structures observed inside the prominences, like well outlined knots of plasma or extended flows, are locally highly dynamic and move along complicated, curved trajectories. Precise measurements of magnetic fields in solar prominences are crucial for understanding their structure, magnetic support, and evolution. Direct investigations of the magnetic fields in prominences started in the early '60s, using at first the Zeeman effect \cite{ZS1961,R1967,TH2011}, subsequently replaced by spectropolarimetry of prominences applying Zeeman and Hanle effect observations \cite{L1977,BSB1978}. The majority of the absolute field strength measurements in quiescent prominences revealed fields of less than 10 Gauss, but some stronger local fields up to 70 Gauss were also reported \cite{Q1985,TH1995,P2001,AD2003,Cae2003,WB2003,M2006,Ch2010,H2010}. It is worth to stress, that Kuckein and co-workers reported recently 600-700 Gauss magnetic fields in a filament (\citeauthor{K2009}, \citeyear{K2009}; see also \citeauthor{J2010}, \citeyear{J2010}), ascribing low values of the earlier measurements to the lack of full Stokes polarimetry.

Surges are active prominences that look like long straight or curved columns of plasma shot out from the solar surface with velocities up to a few hundred km~s$^{-1}$  \cite{G1982}, their typical density could be estimated as roughly $5 \times 10^{-12}$ g cm$^{-3}$ ($n_{e}$=10$^{11}$-10$^{12}$ cm$^{-3}$) \cite{Bru1977,TH1995}. The magnetic fields in active prominences (like surges or sprays) are less known than magnetic fields in quiescent ones \cite{TH2011}. However, in the case of surges there exist some estimations of magnetic field strengths, ranging from a few up to a few tens of Gauss \cite {L2004,Bea2007,SAN2008}.

Direct observations and modeling of the magnetic fields in solar prominences are still difficult while the resolving power of the existing relevant observing facilities is substantially lower than the diameters of prominence subtle structures (\textit{i.e.} threads). Thus, each opportunity for independent check of the obtained results should be explored. In our previous paper \cite{Zapior2010} we presented in detail a method of restoration of the true 3D trajectories of prominence knots using ground-based observations taken with a single telescope equipped with a \textit{Multi-channel Subtractive Double Pass} (MSDP) imaging spectrograph \cite{Mein77}. We have shown that the method allows an evaluation of the true 3D trajectories of the prominence knots without any assumptions concerning the shape of trajectories or dynamics of the motion. In the present work we exploit the method for the determination of accelerations acting on observed plasma knots and for the estimation of lower limits of the magnetic fields channeling the observed plasma flows under the assumption that magnetic field configurations fully control plasma trajectories.

\section{Observations and data processing}

We investigated 13 plasma knots recorded in two surges (\textit{i.e.} active prominences) observed on July 7, 2011 at the west limb of the Sun, two quiescent prominences observed over the north-west solar limb on August 18, 2011 and over the east solar limb on September 3, 2011, respectively, as well as an activated prominence observed over the north-west solar limb on September 1, 2011.

Spectrograms of the investigated prominences were collected with the Large Coronagraph equipped with an MSDP imaging spectrograph at the Bia{\l}k{\'o}w Observatory of the University of Wroc{\l}aw, Poland. The coronagraph has a 51~cm entrance aperture and nearly 14.5 m effective focal length, its effective spatial resolution is limited by seeing only. The MSDP spectrograph has a 9-channel \emph{prism-box} and its spectral resolution is equal to 0.4 \AA{}. The field of view (FOV) of an individual spectrograms is equal to ($325 \times 41 \textrm{ arcsec}^2$) on the Sun. The numerical reduction of the raw spectrorgams was made using the standard MSDP software designed by Mein \cite{Mein91a,Mein91b,Rudawy1995} and our own auxiliary codes. After numerical processing we obtained for each spectrogram a compound file containing 13 quasi-monochromatic images encompassing the whole FOV and separated by $\Delta \lambda=0.2$ \AA{} in wavelengths along the H$\alpha$ line profile and the H$\alpha$ line spectra for all pixels inside the FOV (in a range of $\Delta \lambda= \pm 1.2$ \AA{} from line center). Series of consecutive individual quasi-monochromatic images (up to 25 images per series) are numerically ``combined'' into large-area images covering the whole investigated object (like a prominence). The effective time resolution of the processed data is of the order of 15-60 seconds, depending on the size of the observed object.

The images were co-aligned in a common triaxial reference frame having arbitrary oriented axes: \emph{X} and \emph{Y} axes lie in the sky plane, \emph{Z}-axis is oriented outward the observer. \emph{X-Y} positions of the knot are calculated on the consecutive images using the position of the centroid of the knot delimited by an arbitrary selected isophote (usually the isophote of 70\% of the brightest pixel); the mean emission profiles of the knot are used to evaluate the line of sight (LOS) velocities of the material. Subsequently, the 3D trajectories of the knots are calculated using a polynomial approximation of the obtained positions in the plane of the sky (expressed in kilometers) and LOS velocities (in kilometers per second) at discrete times and then by integration of the velocities the translation along the \emph{Z}-axis is obtained. The absolute starting positions of the knots along the \emph{Z}-axis could not be determined with our method and thus we assumed for each knot $z=0$ for starting or ending position. The detailed description of the applied reduction procedures was discussed in \citeauthor{Zapior2010} (\citeyear{Zapior2010}). 
The present version of the code automatically follows the positions of the selected knots on consecutive images, using simple knot identification and a tracking algorithm, which takes into account instantaneous changes of the brightness, shape and seeing-induced distortions of the knots. The code works completely autonomously after initiation, but due to the high rate of misidentifications, caused mainly by overlapping structures, all the results are examined by the observer to prevent bad tracking. An obvious extension of the code provides also the spatial velocity, the tangential (along the trajectory) and centripetal (perpendicular to the trajectory) accelerations of the observed plasma knots. The accelerations obtained for all investigated prominence knots are described in Section \ref{res}.

Using restored 3D trajectories of the individual knots we estimated the local magnetic fields of the prominences under the assumption that the magnetic field fully controls the observed trajectories of plasma knots.
Supposing the equipartition principle between magnetic and kinetic energy densities and disregarding an influence of gravitation and the viscosity we have 
$$ \frac{1}{2} \rho v^{2}= \frac{B^2}{8\pi}, $$
where $\rho$ is the knot density, $v$ is the instantaneous spatial velocity of the knot and $B$ is the magnetic field for different times during the observation of an individual knot. We made similar consideration as Ballester and Kleczek \cite{Ball84} for the motion of the knots in three dimentions and obtained the expression: 
$$B = \sqrt{4\pi \rho v^2},$$
which is the lower limit of the instantaneous magnetic field of the prominence.
However, while variations of the diameters of magnetic flux ropes channeling the investigated knots are unknown, we present in graphical form field strengths calculated along the whole observed trajectory of each particular knot and in 
Table~\ref{tab:18knots} the highest and lowest values of the fields instead of giving a single minimal value. 

Electron densities of the solar prominences estimated using various observational methods and theoretical models vary greatly (see, for example \citeauthor{Mal1989}, \citeyear{Mal1989}; \citeauthor{TH1995}, \citeyear{TH1995}; \citeauthor{Lab2010}, \citeyear{Lab2010} and references therein). The variations are caused both by differences between the various applied observational techniques and by real variations among individual prominences and various kinds of prominences \cite{Lab2010}. Taking into account a broad range of the presented results, typical electron densities of the prominences could be characterized as 10$^{9}$ - 10$^{11}$ cm$^{-3}$. 
While the estimated lower limits of the magnetic fields controlling prominences are proportional to the square root of the electron densities, we assumed in our calculations high, but reasonable electron densities of the plasma equal to 
 $n_{e} = 5 \times 10^{11} $ cm$^{-3}$ in surges \cite{Bru1977} and 
$n_{e} = 5 \times 10^{10} $ cm$^{-3}$ in quiescent/activated prominences \cite{Lab2010}.

Time spans of the knots' observations, extreme values of the measured accelerations and estimated limiting magnetic fields are presented concisely in Table~\ref{tab:18knots}.
The errors of the measured \emph{X} and \emph{Y} position of the knots (\textit{i.e.} positions measured in the sky plane) were caused mainly by errors of the co-alignment of the consecutive images in the common triaxial reference frame. These errors were estimated as a standard deviation of the measured positions of selected stable objects (like the solar limb in the \emph{X}-direction or selected structures in the prominences in the \emph{Y}-direction). The errors of the \emph{X} and \emph{Y} positions were usually less than 700 km, while the diameters of the investigated knots were of the order of $d=3-4 \times 10^{4}$ km. The errors of the radial velocities were evaluated using the bootstrap method \cite{Ef1979,An2010}. They were mostly of the order of 0.5 km s$^{-1}$, but in cases of very faint, barely visible knots they increased up to 10 km s$^{-1}$. The errors of the \emph{Z} positions were estimated as the accumulation of the individual errors of all previous positions along the trajectory, usually reaching about 800 km per 1000 seconds of observations. Error ranges of the accelerations and lower limits of the magnetic fields were calculated using bootstrap method also; they are plotted by dashed lines on Figures \ref{fig:07g}, \ref{fig:18g}, \ref{fig:01g} and  \ref{fig:03g}.

\section{Results} \label{res}

\subsection{Surges on July 7, 2011}

\begin{figure}[t]
\vspace{0.5cm}
\centerline{\includegraphics[width=0.6\textwidth,clip]{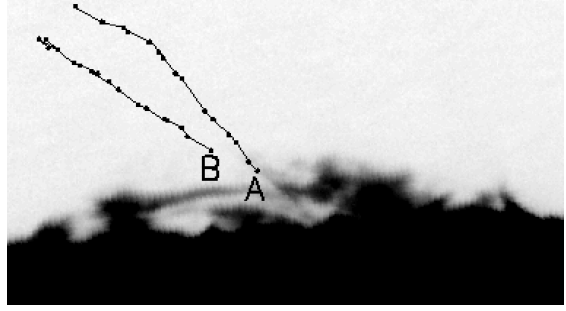}}
\caption{Temporal changes of the positions projected on the sky plane of the two knots forming the tips of two surges (marked A and B) observed on July 7, 2011. The positions are superimposed on a ``combined'' image (see main text for details) taken during an early phase of the surge marked \emph{A} at 15:58~UT in H$\alpha$ line center.}\label{fig:072d}
\end{figure}

\begin{figure}[h]
\vspace{0.5cm}
\centerline{\includegraphics[width=0.7 \textwidth,clip]{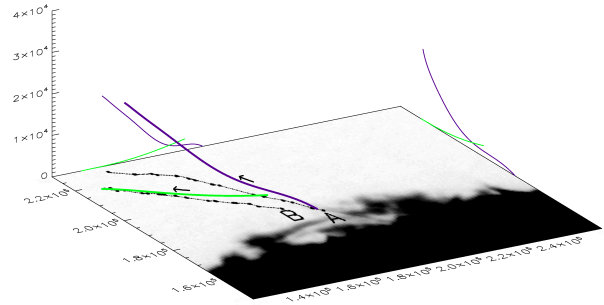}}
\caption{Restored 3D trajectories of the tips of the two surges observed on July 7, 2011. Axes are scaled in
kilometers. Thick lines represent axonometric projections of the trajectories while thin lines represent projections of the trajectories on the perpendicular surfaces of the reference frame X-Z and Y-Z. LOS is toward  the top (along the Z axis). Purple lines represent surge A, green lines represent surge B. Small black arrows show directions of motions of the investigated knots along the trajectories.}\label{fig:073d}
\end{figure}

Two surges were observed on July 7, 2011 over the west limb of the Sun, near NOAA 11244 active region (AR) (PA=$285{^\circ}$-$300{^\circ}$). The observational data were collected between 05:22 UT and 17:06 UT with time cadence between 18 and 45 seconds, the total number of the processed multi-wavelength compound images is 766. Figure \ref{fig:072d} presents a ``combined'' image taken during an early phase of the surge marked \emph{A} at 15:58~UT in H$\alpha$ line center with superimposed measured positions in the sky plane of the two knots forming the tips of both surges. Figure \ref{fig:073d} presents the restored 3D trajectories of both knots. For better visibility of both tracks we the shifted starting positions of the knots by approximately 5000 km in the vertical direction.

\begin{figure} [h]
\centerline{\includegraphics[width=0.9\textwidth,clip]{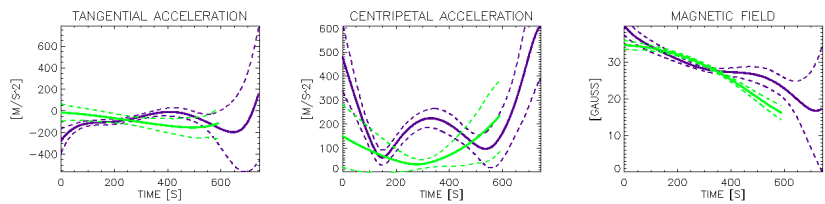}}
\caption{Tangential accelerations (left), centripetal accelerations (center) and calculated lower limits of magnetic field strengths along restored trajectories of the two surges on July 7, 2011. For each knot time is counted from zero from the beginning of its observations. Purple lines represent surge A, green lines represent surge B (compare to Figure \ref{fig:073d}). Error ranges for accelerations and estimated lower limits
of the magnetic fields are plotted by dashed lines.}\label{fig:07g}
\end{figure}

Knot A decelerated from $\sim120$ ~km s$^{-1}$ to $\sim60$~km s$^{-1}$, knot B had a nearly constant velocity of $\sim90$~km/s. However, the LOS component of the velocity was negative for knot A and positive for knot B. Thus their trajectories observed in projection on the sky plane (2D) looks nearly parallel, while in 3D they disperse toward and outward the observer. The tangential accelerations measured for both knots were negative, ranging from -150 m s$^{-2}$ to 0 m s$^{-2}$. The centripetal accelerations varied between 50 m s$^{-2}$ and 250 m s$^{-2}$ (excluding the starting and especially the ending parts of the observational period of \emph{A} knot, when the calculated accelerations and magnetic fields are affected by side-effects of the polynomial approximation for the trajectory, which significantly increased estimated error ranges), see Figure \ref{fig:07g}, left and middle panels. The values strongly affected by side-effects are marked with an asterisk in Table \ref{tab:18knots}.

The minimum magnetic fields calculated along the trajectory of knot \emph{A} range between 18 and 40 Gauss, while for knot \emph{B} they range from 16 to 34 Gauss (Figure \ref{fig:07g}, right panel).

\subsection{The activated prominence on August 18, 2011}

\begin{figure} [t]
\centerline{\includegraphics[width=0.7\textwidth,clip]{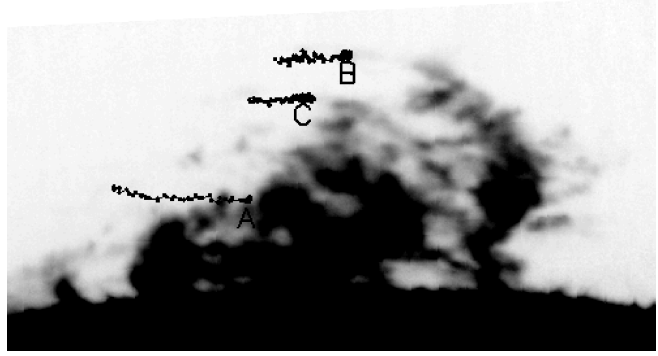}}
\caption{Temporal changes of the positions projected on the sky plane of three plasma knots (marked A, B and C) observed in the activated prominence recorded over the north-west solar limb on August 18, 2011. The positions are superimposed on a ``combined'' image taken at 10:41~UT in H$\alpha$ line center.}\label{fig:182d}
\end{figure}

\begin{figure} [h]
\centerline{\includegraphics[width=0.95\textwidth,clip]{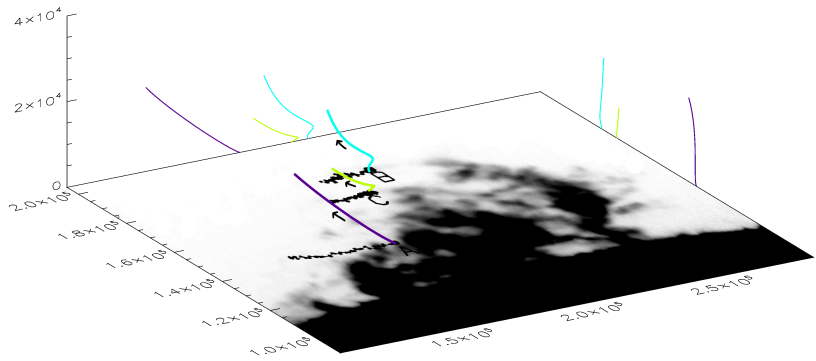}}
\caption{Restored 3D trajectories of the three knots observed in the activated prominence on August 18, 2011. Trajectories are presented as well as axes are scaled and oriented as in Figure 2. Purple, blue and green lines represent knots A, B and C, respectively.}\label{fig:183d}
\end{figure}

The activated prominence was observed on August 18, 2011, from 08:27 UT up to 16:16 UT, over the north-west part of the solar limb (PA=$315{^\circ}$-$335{^\circ}$). The time cadence of the data was 30-45 seconds, the total number of the collected multi-wavelength compound images of the prominence was 714. Figure~\ref{fig:182d} presents a ``combined'' (2D) image of the prominence taken at 10:41 UT in H$\alpha$ line center with superimposed positions measured in the sky plane of three investigated knots (marked \emph{A}, \emph{B} and \emph{C}, respectively). Apparent 2D motions of knots were roughly parallel to the limb. Figure \ref{fig:183d} presents restored 3D trajectories of the same knots. The knots were located in different parts of the prominence, but all moved toward the observer, having quite similar, nearly parallel trajectories, inclined by $\sim45{^\circ}$ to the plane of the sky. The main body of the prominence, crossing the central meridian on August 11, 2011, was inclined by $\sim30{^\circ}$ to the local meridian, thus the trajectories of the knots were inclined by $\sim75{^\circ}$ to the prominence's body.

The velocities of all knots where in the range of 5-33 km s$^{-1}$ for knot \emph{A}, 3-21 km s$^{-1}$ for knot \emph{B} and 0-25 km s$^{-1}$ for knot \emph{C}. The tangential accelerations measured for all three knots were small, ranging from -30 m s$^{-2}$ to 52 m s$^{-2}$. Knot \emph{A} at first was accelerated tangentially but with was gradually decreasing acceleration from 52 m s$^{-2}$ to 0 m s$^{-2}$, next it was decelerated with increasing deceleration from 0 m s$^{-2}$ to -30 m s$^{-2}$. Knot \emph{B} showed alternately accelerations and decelerations along the trajectory, ranging from -17 m s$^{-2}$ to 17 m s$^{-2}$ (3D trajectory consisted of three arcs).
Knot \emph{C} at first was slightly decelerated tangentially by $\sim$-15 m s$^{-2}$ and next, after crossing the tip of the trajectory, it was accelerated by $\sim$15 to 18 m s$^{-2}$.

\begin{figure} [h]
\centerline{\includegraphics[width=0.9\textwidth,clip]{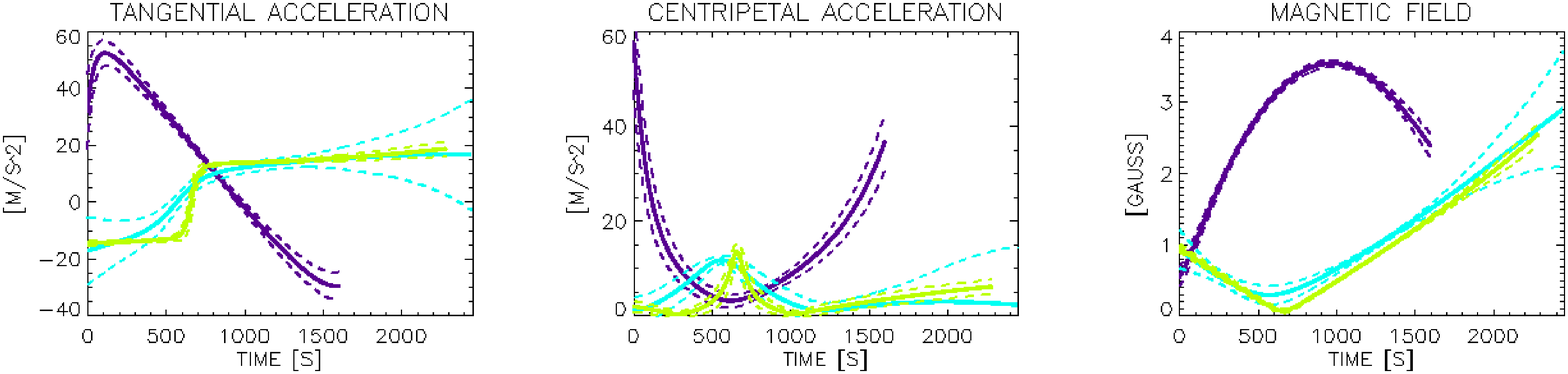}}
\caption{Tangential accelerations (left), centripetal accelerations (center) and calculated lower limits of magnetic field strengths along the restored trajectories of A, B and C knots observed in the activated prominence on August 18, 2011. For each knot time is counted from zero from the beginning of its observations. Color coding is the same as in Figure \ref{fig:183d}. Error ranges for accelerations and estimated lower limits
of the magnetic fields are plotted by dashed lines.}\label{fig:18g}
\end{figure}

The centripetal accelerations of knot \emph{A} changed smoothly due to the very regular shape of the trajectory, ranging between 3 and 57 m s$^{-2}$. The centripetal accelerations of knot \emph{B} were more variable in time (again, due to the complicated trajectory of the knot), ranging from 1 to 12 m s$^{-2}$. Similarly, the centripetal accelerations of knot \emph{C} ranged from 0 to 13 m s$^{-2}$ (see Figure \ref{fig:18g}, left and middle panels).

The estimated minimum magnetic fields calculated along the trajectory of knot \emph{A} range between 0.5 and 3.5 Gauss, while for knots \emph{B} and \emph{C} they range from 0.2 to 2.9 and 0.0 to 2.5 Gauss respectively (Figure \ref{fig:18g}, right panel).

\subsection{The activated prominence on September 1, 2011}

\begin{figure} [t]
\centerline{\includegraphics[width=0.6\textwidth,clip]{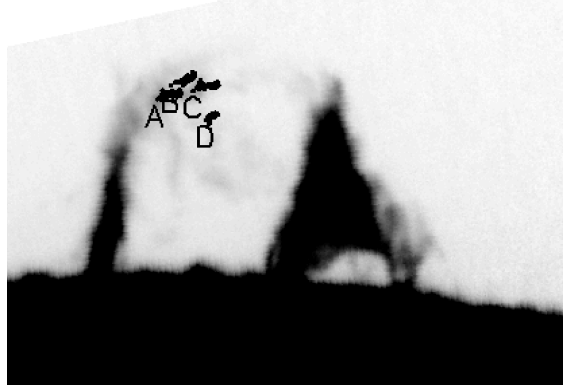}}
\caption{Temporal changes of the positions projected on the sky plane of four plasma knots (marked A, B, C and D) observed in the activated prominence recorded over the north-west solar limb on September 1, 2011. The positions are superimposed on a ``combined'' image taken at 13:09~UT in H$\alpha$ line center.}\label{fig:012d}
\end{figure}

\begin{figure} [h]
\centerline{\includegraphics[width=0.8\textwidth,clip]{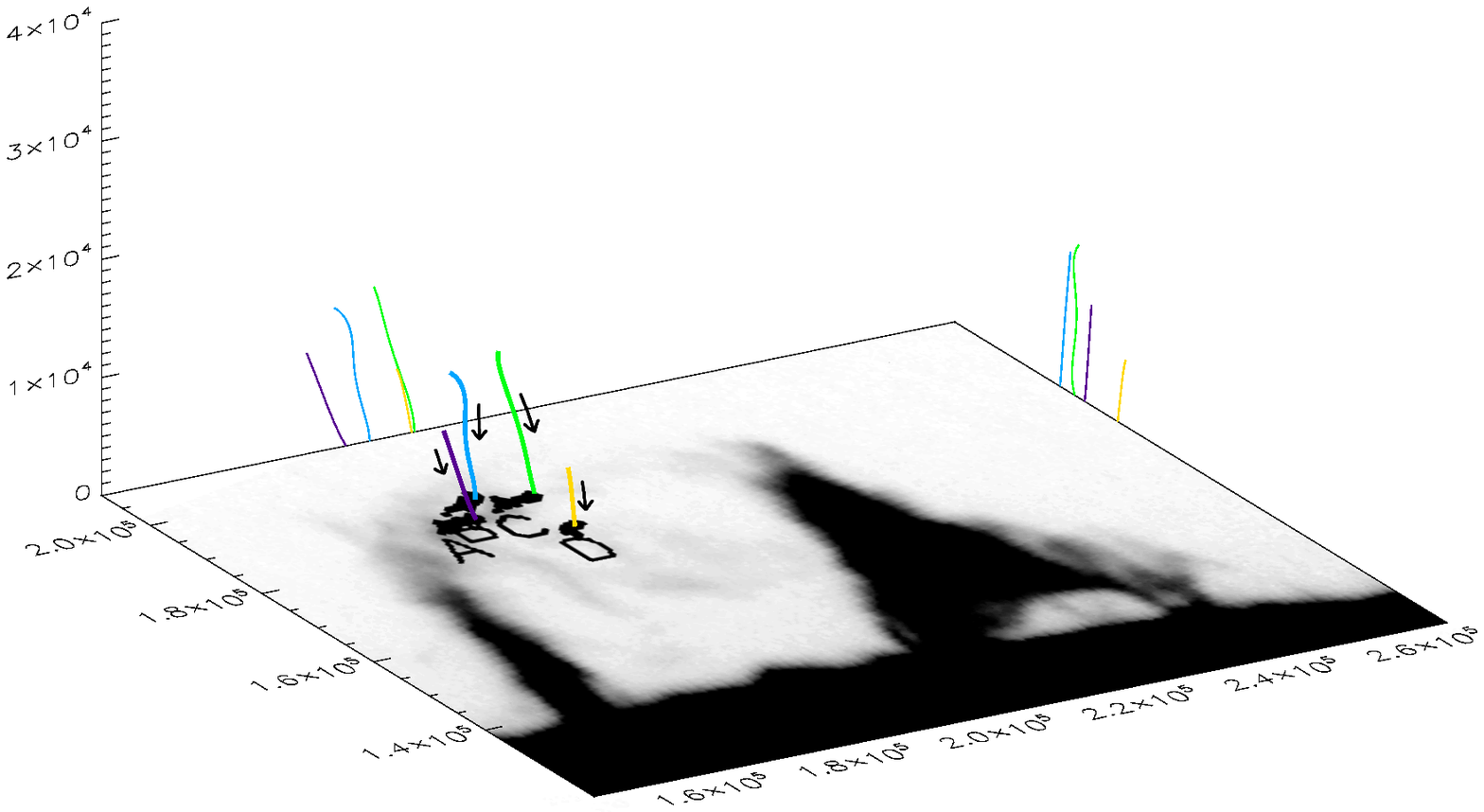}}
\caption{Restored 3D trajectories of four knots observed in the activated prominence on September 1, 2011. Trajectories are presented as well as axes are scaled and oriented as in Figure 2. Purple, blue, green and yellow lines represent knots A, B, C, and D respectively.}\label{fig:013d}
\end{figure}

The activated prominence was visible over the north-west solar limb on September 1, 2011 (PA=$330{^\circ}$). The observations began at 08:00 UT and stopped at 14:24 UT, the applied time cadence varied between 36 and 60 seconds. The total number of collected images was equal to 410. Figure \ref{fig:012d} and Figure \ref{fig:013d} present an image of the prominence taken at 13:09 UT in H$\alpha$ line center with over-plotted measured consecutive positions of four knots (marked A, B, C and D) in the sky plane and restored positions in 3D space, respectively. Apparent 2D displacements of all knots in the sky plane were small, but when observed in 3D, all knots moved nearly perpendicularly to the plane of the sky and nearly parallel to each other along slightly bent trajectories, having similar velocities (of the order of 10 km s$^{-1}$). It is worth to stress, that knots were observed in various periods of time (\emph{A} and \emph{C} between 13:09 UT and 13:32 UT, \emph{B} between 13:42 UT and 14:06 UT and \emph{D} between 11:02 UT and 11:15 UT).

\begin{figure} [h]
\centerline{\includegraphics[width=0.9\textwidth,clip]{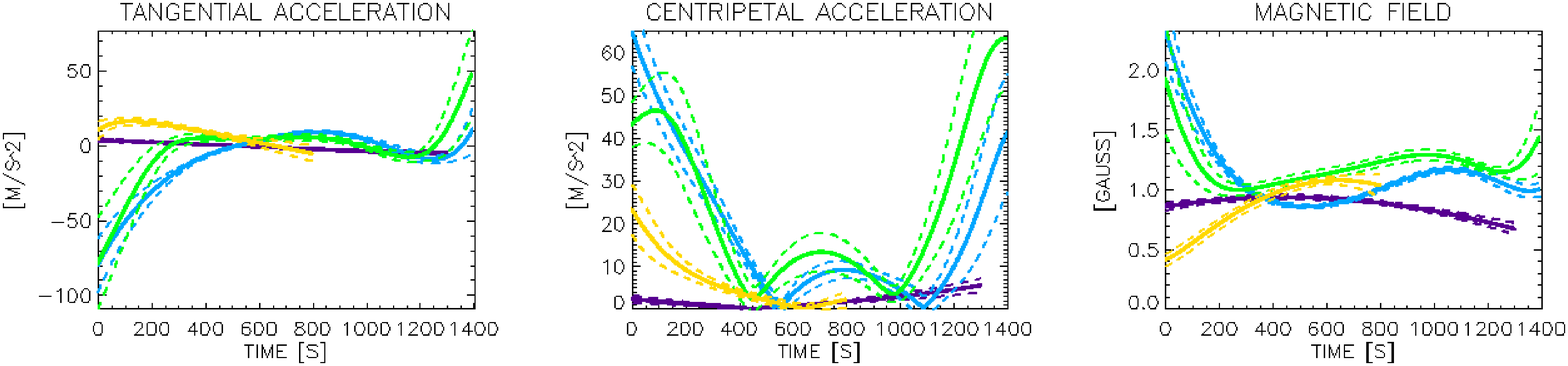}}
\caption{Tangential accelerations (left), centripetal accelerations (center) and calculated lower limits of magnetic field strengths along the restored trajectories of the knots observed in the activated prominence on September 1, 2011. For each knot time is counted from zero from the beginning of its observations. Color coding is the same as in Figure \ref{fig:013d}. Error ranges for accelerations and estimated lower limits
of the magnetic fields are plotted by dashed lines.}\label{fig:01g}
\end{figure}

The velocities of all knots where in the range 4 - 23 km s$^{-1}$.
During most of the observing periods their tangential accelerations were small, in a range from -20 m s$^{-2}$ to 20 m s$^{-2}$. The centripetal acceleration ranged from 0~m~s$^{-2}$ to 25~m~s$^{-2}$. High tangential and centripetal accelerations of knots \emph{B} and \emph{C} estimated at the beginning and end of the observed periods seem to be artifacts caused by side-effects of the polynomial approximation to their trajectories (see Figure \ref{fig:01g}, left and middle panels).

The estimated minimum magnetic fields calculated along the trajectories of all knots were about 1 Gauss (excluding side-effects of the polynomial approximation to the trajectories knots \emph{B} and \emph{C}), see Figure \ref{fig:01g}, right panel.

\subsection{The quiescent prominence on September 3, 2011}

\begin{figure} [t]
\centerline{\includegraphics[width=0.7\textwidth,clip]{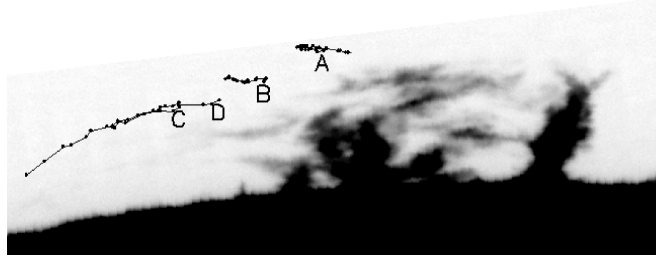}}
\caption{Temporal changes of the positions projected on the sky plane of four plasma knots (marked A, B, C and D) observed in the quiescent prominence recorded over the east solar limb on September 3, 2011. The positions are superimposed on a ``combined'' image taken at 09:16~UT in H$\alpha$ line center.}\label{fig:032d}
\end{figure}

\begin{figure} [h]
\centerline{\includegraphics[width=0.95\textwidth,clip]{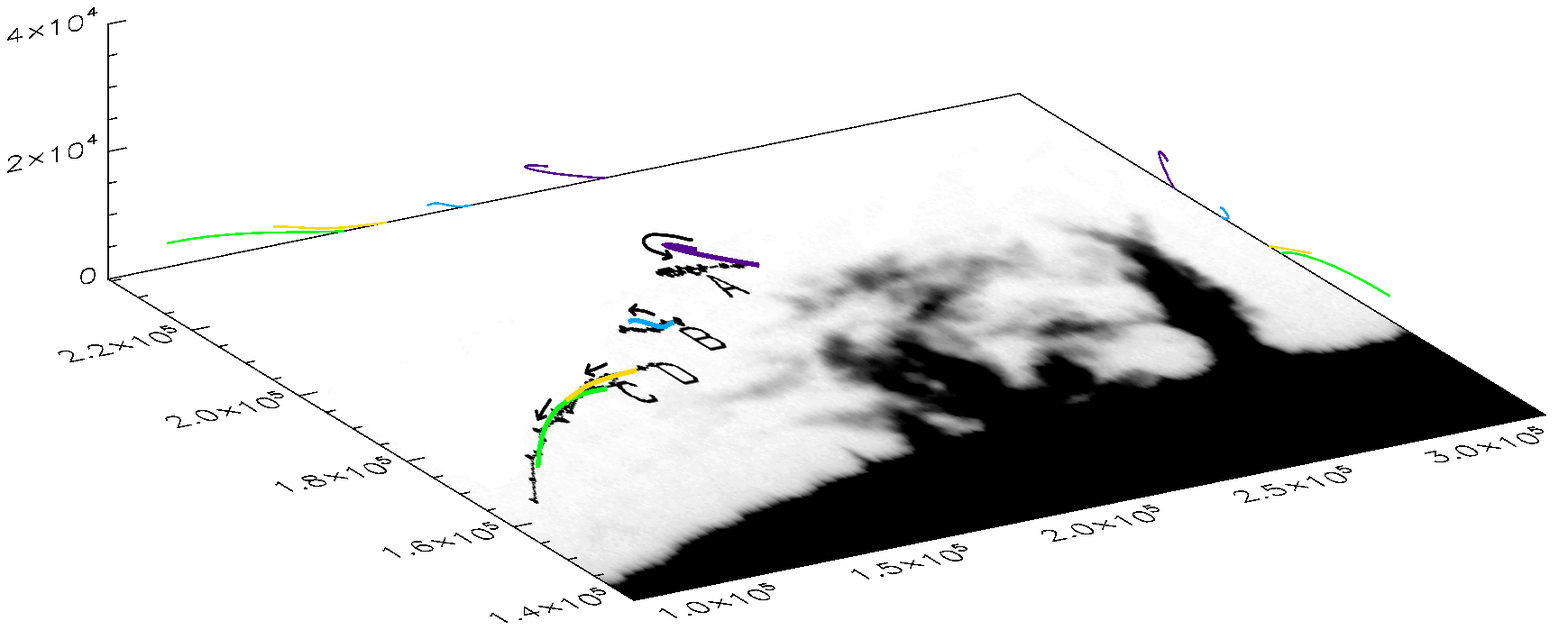}}
\caption{Restored 3D trajectories of four knots observed in the quiescent prominence on September 3, 2011. Trajectories are presented as well as axes are scaled and oriented as in Figure 2. Purple, blue, green, and yellow lines represent knots A, B, C and D, respectively.}\label{fig:033d}
\end{figure}

The quiescent prominence was observed over the east solar limb on September~3, 2011 (PA=$55{^\circ}$-$70{^\circ}$). The observations started at 09:14 UT and finished at 15:17 UT, time cadence was between 60 and 72 seconds. The total number of collected images was 299. Figure \ref{fig:032d} presents an image of the prominence taken at 09:16 UT in H$\alpha$ line center with over-plotted measured consecutive positions in the sky plane of the four investigated knots (marked \emph{A}, \emph{B}, \emph{C} and \emph{D}, respectively). Apparent 2D motions of all knots were directed toward the south and they outlined slightly curved loops. Figure \ref{fig:033d} presents their restored 3D trajectories. LOS velocities of all knots were very small so their true 3D trajectories were similar to the trajectories projected on the sky plane. All knots moved toward the observer, along little inclined, curved trajectories, except knot \emph{A} which turned back.

The velocities of all knots where in the range of 6-8 km s$^{-1}$ for knot \emph{A}, 9-23~km~s$^{-1}$ for knot \emph{B}, 10-19 km s$^{-1}$ for knot \emph{C} and 4-10 km s$^{-1}$ for knot \emph{D}.

\begin{figure} [h]
\centerline{\includegraphics[width=0.9\textwidth,clip]{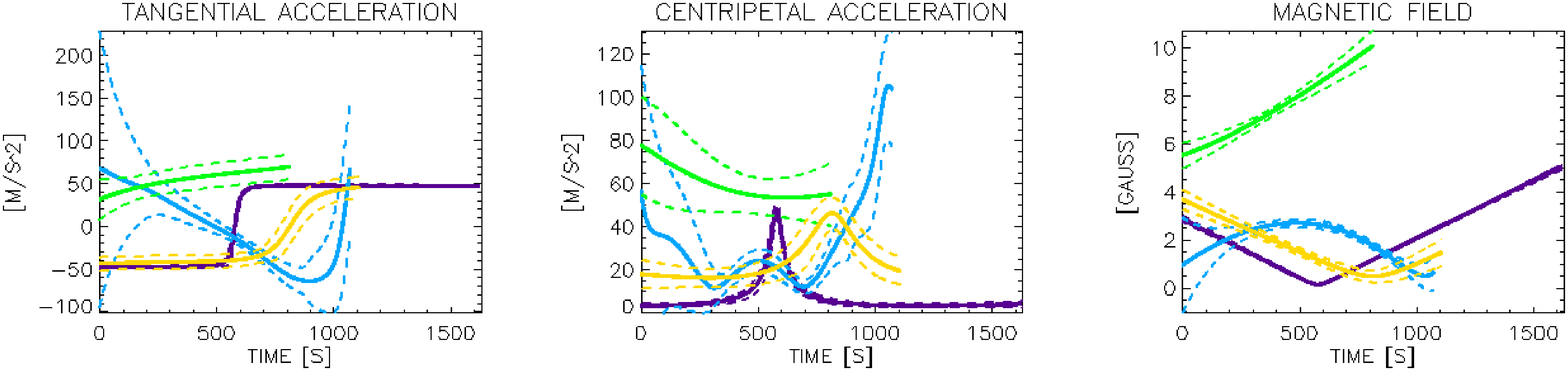}}
\caption{Tangential accelerations (left), centripetal accelerations (center) and calculated lower limits of magnetic field strengths along the restored trajectories of the knots observed in the quiescent prominence on September 3, 2011. For each knot time is counted from zero from the beginning of its observation. Color coding is the same as in Figure \ref{fig:033d}. Error ranges for accelerations and estimated lower limits
of the magnetic fields are plotted by dashed lines.}\label{fig:03g}
\end{figure}

The tangential accelerations measured for all knots were relatively small, ranging from -70 m s$^{-2}$ to 60 m s$^{-2}$. Knots \emph{A} and \emph{D} were firstly decelerated by -50 m s$^{-2}$ and next accelerated by 50 m s$^{-2}$ (they were observed in various time intervals). Knot \emph{B} at first accelerated with the acceleration decreasing gradually from 60 m s$^{-2}$ down to 0 m s$^{-2}$, next it decelerated with deceleration growing from 0 m s$^{-2}$ up to -60 m s$^{-2}$. Rapid changes of tangential and centripetal accelerations of knot \emph{B} at the end of the observed period seems to be an artifact caused by side-effects of the polynomial approximation to its trajectory. Knot~\emph{C} was accelerated along the whole observed trajectory, with an acceleration growing from 35 m s$^{-2}$ to 70 m s$^{-2}$. The centripetal accelerations of the knots ranged from 5 to 80 m s$^{-2}$ (see Figure \ref{fig:03g}, left and middle panels).

The estimated minimum magnetic fields calculated along the trajectory of knot \emph{C} range between 5 and 10 Gauss, while for knots \emph{A}, \emph{B} and \emph{D} they range from 0.1 to 5 Gauss (Figure \ref{fig:03g}, right panel).

\section{Conclusions}

We have presented 3D trajectories, velocities, tangential and centripetal accelerations of the plasma knots observed in solar prominences of various kinds (surges, activated, and quiescent prominences), restored using single ground-based telescope and MSDP spectrograph observations. 
The established parameters of the 3D (spatial) motions of the individual knots were applied to estimate lower limits of the magnetic fields channeling observed plasma flows under the assumption of the equipartition principle.
The applied assumption is very strong, but it is consistent with contemporary prominence models assuming that local magnetic fields guide plasma flows and determine their geometry, being revealed by trajectories of the visible individual plasma knots or streams \cite{AD2003}. We found also that the lower limits of the magnetic fields varied along the trajectories (\textit{i.e.} along the magnetic ropes). However, while the diameter variations of the magnetic ropes channeling investigated knots are unknown, no unique values of the magnetic field were calculated for the whole trajectories, but instant values along the trajectories only.

The tangential accelerations of the knots observed in active and quiescent prominences ranged roughly from -100 m s$^{-2}$ to 80 m s$^{-2}$, while in the case of two surges they range from -250 m s$^{-2}$ to -20 m s$^{-2}$. The centripetal accelerations of the knots observed in active and quiescent prominences ranged roughly from 0~m~s$^{-2}$ to 100 m s$^{-2}$, while for the two surges they ranged from 30 m s$^{-2}$ to $\sim250$ m s$^{-2}$. The measured accelerations are roughly similar to previously presented results \cite{KP1985,R1990,V1990,A2012}.

Assuming approximate typical electron densities of the plasma $n_e = 5\times 10^{11}$~cm$^{-3}$ in surges and  $n_e = 5\times 10^{10}$ cm$^{-3}$ in quiescent/activated prominences \cite{Bru1977,Lab2010} we found that lower limits of the local magnetic fields in surges were equal to 16-40 Gauss, while in activated and quiescent prominences were less than 10 Gauss. In case if actual densities of the investigated prominences were smaller than assumed typical values applied in calculations, the obtained lower limits of the magnetic fields controlling prominences are slightly overestimated.
Most of the direct measurements of the solar prominence magnetic fields, made using various methods also revealed rather weak local magnetic fields, ranging from a few up to a few tens of Gauss (see the discussion in the Introduction and references therein). However, as it was mentioned before, there were also some recent observations and numerical models reporting magnetic field strengths in the filaments of the order of 600-700 Gauss \cite{K2009,J2010}. Our result is consistent with previous detections of weak local magnetic fields in solar prominences, but, establishing the lower limits of the magnetic fields for a limited number of prominence knots only, we cannot discard a possibility of strong fields in prominences.

\begin{acks}
MZ is supported by Human Capital - European Social Fund.
\end{acks}

\bibliographystyle{spr-mp-sola}

\begin{thebibliography}{}

\bibitem[\protect\citeauthoryear{{Andrae}}{2010}]{An2010}Andrae,~R.: 2010, arXiv:1009.2755v3
\bibitem[\protect\citeauthoryear{{Antolin and Rouppe van der Voort}}{2012}]{A2012}Antolin, P., Rouppe van der Voort, L.: 2012, \apj, \textbf{745}, 152
\bibitem[\protect\citeauthoryear{{Aulanier and Demoulin}}{2003}]{AD2003}Aulanier,~G., Demoulin,~P.: 2003, \aap{} \textbf{402}, 769
\bibitem[\protect\citeauthoryear{{Ballegooijen and Cranmer}}{2010}]{BC2010}Ballegooijen,~A.A, Cranmer,~S.R.: 2010, \apj{} \textbf{711}, 164
\bibitem[\protect\citeauthoryear{{Ballester and Kleczek}}{1984}]{Ball84}Ballester,~J.~L., Kleczek,~J.: 1984, \solphys{} \textbf{90}, 37
\bibitem[\protect\citeauthoryear{{Bommier and Sahal-Brechot}}{1978}]{BSB1978}Bommier,~V., Sahal-Brechot,~S.: 1978, \aap{} \textbf{69}, 57
\bibitem[\protect\citeauthoryear{{Brooks, Kurokawa and Berger}}{2007}]{Bea2007}Brooks,~D.H., Kurokawa,~H., Berger,~T.E.: 2007, \apj{} \textbf{656}, 1197
\bibitem[\protect\citeauthoryear{{Casini \textit{et al.}}}{2003}]{Cae2003}Casini,~R., L\'opez Ariste,~A., Tomczyk,~S., Lites,~B.W.: 2003, \apj{} \textbf{598}, 67
\bibitem[\protect\citeauthoryear{{Chae}}{2010}]{Ch2010}Chae,~J.: 2010, \apj{} \textbf{714}, 618
\bibitem[\protect\citeauthoryear{{Efron}}{1979}]{Ef1979}Efron,~B.: 1979, \textit{Ann. Statist.}, \textbf{7}, 1
\bibitem[\protect\citeauthoryear{{Garczy{\'n}ska \textit{et al.}}}{1982}]{G1982}Garczy{\'n}ska,~N., Rompolt,~B., Benz,~A.O., Slottje,~C., Tlamicha,~A., Zanelli,~C.: 1982, \solphys{} \textbf{77}, 277
\bibitem[\protect\citeauthoryear{{Hiller, Shibata and Isobe}}{2010}]{H2010}Hillier,~A., Shibata,~K., Isobe,~H.: 2010, \emph{PASJ} \textbf{62}, 1231
\bibitem[\protect\citeauthoryear{{Jing \textit{et al.}}}{2010}]{J2010}Jing,~J., Yuan,~Y., Wiegelmann,~T., Xu,~Y., Liu,~R., Wang,~H.: 2010, \apj{} \textbf{719}, L56
\bibitem[\protect\citeauthoryear{{Klepikov and Platov}}{1985}]{KP1985}Klepikov,~V.Y., Platov,~Y.V.: 1985, \emph{Astronom. Zhurnal} \textbf{62}, 983
\bibitem[\protect\citeauthoryear{{Kuckein \textit{et al.}}}{2009}]{K2009}Kuckein,~C., Centeno,~R., Mart\'inez Pillet,~V., Casini,~R., Manso Sainz,~R., Shimizu,~T.: 2009, \aap{} \textbf{501}, 1113
\bibitem[\protect\citeauthoryear{{Labrosse \textit{et al}.}}{2010}]{Lab2010} Labrosse, N., Heinzel, P., Vial, J.-C., Kucera, T., Parenti, S., Gun\'ar, Schmieder, B., Kilper, G.: 2010, \textit{Space Sci. Rev.} \textbf{151}, 243
\bibitem[\protect\citeauthoryear{{Liu and Kurokawa}}{2004}]{L2004}Liu,~Y., Kurokawa,~H.: 2004, \apj{} \textbf{610}, 1136
\bibitem[\protect\citeauthoryear{{Leroy, Ratier and Bommier}}{1977}]{L1977}Leroy,~J.L., Ratier,~G., Bommier,~V.: 1977, \aap{} \textbf{54}, 811
\bibitem[\protect\citeauthoryear{{Malherbe}}{1977}]{Mal1989} Malherbe, J.-M.: 1989, In: Priest, E.~R. (ed.) \textit{Dynamics and Structure of Quiescent Solar Prominences, Astrophysics and Space Science Library}, \textbf{150}, 115
\bibitem[\protect\citeauthoryear{{Mein}}{1977}]{Mein77} Mein,~P.: 1977, \solphys{} \textbf{54}, 45
\bibitem[\protect\citeauthoryear{{Mein}}{1991a}]{Mein91a} Mein,~P.: 1991a, \aap{} \textbf{248}, 237
\bibitem[\protect\citeauthoryear{{Mein}}{1991b}]{Mein91b} Mein,~P.: 1991b, \aap{} \textbf{248}, 669
\bibitem[\protect\citeauthoryear{{Merenda \textit{et al.}}}{2006}]{M2006}Merenda,~L., Trujillo Bueno,~J., Landi Degl'innocenti,~E., Collados,~M.: 2006, \apj{} \textbf{642}, 554
\bibitem[\protect\citeauthoryear{{Paletou \textit{et al.}}}{2001}]{P2001}Paletou,~F., Lopez Ariste,~A., Bommier,~V., Semel,~M.: 2001, \aap{} \textbf{375}, L39
\bibitem[\protect\citeauthoryear{{Querfeld \textit{et al.}}}{1985}]{Q1985}Querfeld,~C.W., Smartt,~R.N., Bommier,~V., Landi Degl'innocenti,~E.: 1985, \solphys{} \textbf{96}, 277
\bibitem[\protect\citeauthoryear{{Rompolt}}{1990}]{R1990}Rompolt,~B.: 1990, \emph{Hvar Obs. Bull.} \textbf{14}, 37
\bibitem[\protect\citeauthoryear{{Rudawy}}{1996}]{Rudawy1995} Rudawy,~P.: 1996, {\it JOSO Ann. Rep.}, 159
\bibitem[\protect\citeauthoryear{{Rust}}{1967}]{R1967}Rust,~D.M.: 1967, \emph{Astrophys. J.} \textbf{150}, 313
\bibitem[\protect\citeauthoryear{{S\'anchez-Andrade Nu{\~n}o \textit{et al.}}}{2008}]{SAN2008}S\'anchez-Andrade Nu\~no,~B., Bello Gonz\'alez,~N., Blanco Rodr\'iguez,~J., Kneer,~F., Puschmann,~K. G.: 2008, \aap{} \textbf{486}, 577
\bibitem[\protect\citeauthoryear{{Slater} and {Aschwanden}}{2007}]{Slater2007} Slater,~G.L., Aschwanden,~M.J.:
    2007, American Geophysical Union, Fall Meeting, abstract \# SH32A-0771
\bibitem[\protect\citeauthoryear{{Tandberg-Hanssen}}{1977}]{Bru1977} Tandberg-Hanssen, E.: 1977, In: Bruzek, A. and Durrant (eds.), C.~J.: \textit{Illustrated glossary for solar and solar-terrestrial physics, Astrophysics and Space Science Library}, \textbf{69}, 97
\bibitem[\protect\citeauthoryear{{Tandberg-Hanssen}}{1995}]{TH1995}Tandberg-Hanssen,~E.: 1995, \emph{The Nature of Solar Prominences}, Kluwer Acad. Publ., Dordrecht, 92
\bibitem[\protect\citeauthoryear{{Tandberg-Hanssen}}{2011}]{TH2011}Tandberg-Hanssen,~E.: 2011, \solphys{} \textbf{269}, 237
\bibitem[\protect\citeauthoryear{{Vr{\v s}nak}}{1990}]{V1990}Vr{\v s}nak,~B.: 1990, \solphys{} \textbf{127}, 129
\bibitem[\protect\citeauthoryear{{Wiehr and Bianda}}{2003}]{WB2003}Wiehr,~E., Bianda,~M.: 2003, \aap{} \textbf{404}, L25
\bibitem[\protect\citeauthoryear{{Wiik, Dere and Schmieder}}{1993}]{W1993}Wiik,~J.E., Dere,~K., Schmieder,~B.: 1993, \aap{} \textbf{273}, 267
\bibitem[\protect\citeauthoryear{{Zapi{\'o}r} and {Rudawy}}{2010}]{Zapior2010} Zapi{\'o}r,~M., Rudawy,~P.: 2010,
    \solphys{} \textbf{267}, 95
\bibitem[\protect\citeauthoryear{{Zirin and Severny}}{1961}]{ZS1961}Zirin,~H., Severny,~A.B.: 1961, \emph{Observatory} \textbf{81}, 155



\end{thebibliography}

\newpage

\newpage

\newpage

\newpage

\newpage

\begin{table}[h]
\begin{tabular}{cccccccc} \hline
Promi- & Knot & \multicolumn{2}{c}{Observations}  & No. of   & \multicolumn{2}{c}{Acceleration [$\textrm{ms}^{-2}$]} & Magnetic \\
     nence             &        &  Begin & End & images &  Tangential & Centripetal    & Field [G] \\ \hline
Jul 7,& A & 15:55:58 & 16:08:23 & 17 & -280$^{*}$ -- 163$^{*}$ & 59 -- 609$^{*}$  & 18 -- 40\\
      2011              & B & 16:47:42 & 16:57:34 & 17 & -155 -- -18 & 31 -- 234 & 16 -- 34 \\
\hline
Aug 18, & A & 13:38:19 & 14:05:03 & 43 & -30 -- 52 & 3 -- 57 & 0.5 -- 3.5\\
2011 & B & 10:40:48 & 11:21:40 & 83 & -17 -- 17 &    1 -- 12 & 0.2 -- 2.9 \\
        & C & 11:30:53 & 12:08:57 & 86 & -15 -- 18 & 0.1 -- 13 & 0.0 -- 2.5 \\
 \hline
Sep 1, & A & 13:09:14 & 13:32:28 & 38 & -5 -- 3 & 0.0 -- 5 &  0.6 -- 0.9 \\
  2011& B & 13:42:57 & 14:06:19 & 43 & -80$^{*}$ -- 12$^{*}$ & 0.3 -- 65$^{*}$ & 0.8 -- 2.3 \\
 & C & 13:09:14 & 13:32:28 & 38 & -79$^{*}$ -- 47$^{*}$ & 2 -- 63$^{*}$ &  0.9 -- 1.9 \\
 & D & 11:02:17 & 11:15:36 & 22 & -5.4 -- 16 & 0 -- 24 & 0.4 -- 1.0 \\
 \hline
Sep 3, & A & 09:16:18 & 09:43:29 & 33 & -48 -- 47 & 3 -- 49 &  0.1 -- 5 \\
  2011 & B & 12:03:41 & 12:17:51 & 16 & -64 -- 67 & 11 -- 105$^{*}$&  0.4 -- 2.6 \\
 & C & 13:26:32 & 13:40:05 & 14 & 31 -- 69 & 53 -- 78 & 5 -- 10 \\
 & D & 14:23:09 & 14:41:37 & 14 & -44 -- 45 & 16 -- 46 & 0.4 -- 4 \\

\hline
\end{tabular}
$^{*}$ Value potentially affected by side-effects of the polynomial approximation to the trajectory.
\caption{Time spans of the observations, number of ``combined'' images, extreme values of the measured accelerations and estimated minimal magnetic fields.}
\label{tab:18knots}
\end{table}

\end{article}
\end{document}